\documentclass{emulateapj}
\usepackage{apjfonts}

\newcommand{\eqref}[1]{Equation (\ref{#1})}

\newcommand{\MESA}{\texttt{MESA}}

\shorttitle{Helium Detonations on WD}

\begin{document}

%\submitted{Accepted to the Astrophysical Journal June 25, 2009}
\title{Laterally Propagating Detonations in Thin Helium Layers on Accreting
White Dwarfs}

\author{
Dean M. Townsley\altaffilmark{1},
Kevin Moore\altaffilmark{2},
and
Lars Bildsten\altaffilmark{2,3},
}

\altaffiltext{1}{
Department of Physics and Astronomy,
University of Alabama, Tuscaloosa, AL; Dean.M.Townsley@ua.edu
}
\altaffiltext{2}{
Department of Physics,
University of California, Santa Barbara, CA
}
\altaffiltext{3}{
Kavli Institute for Theoretical Physics, University of California, Santa
Barbara, CA
}
\begin{abstract}

Theoretical work has shown that intermediate  mass ($0.01M_\odot<M_{\rm
He}<0.1M_\odot$) Helium shells will unstably ignite on the accreting white
dwarf (WD) in an AM CVn binary. For more massive ($M>0.8M_\odot$) WDs, these
helium shells can be dense enough ($>5\times 10^5 \ {\rm g \ cm^{-3}}$) that
the convectively burning region runs away on a timescale comparable to the
sound travel time across the shell; raising the possibility for an explosive
outcome rather than an Eddington limited helium novae. The nature of the
explosion (i.e. deflagration or detonation) remains ambiguous, is certainly
density dependent, and likely breaks spherical symmetry. In the case of
detonation, this causes a laterally propagating front whose properties in
these geometrically thin and low density shells we begin to study here.  Our
calculations show that the radial expansion time of $<0.1$ s leads to
incomplete helium burning, in agreement with recent work by Sim and
collaborators, but that the nuclear energy released is still adequate to
realize a self-sustaining laterally propagating detonation.  These
detonations are slower than the Chapman-Jouguet speed of $1.5\times 10^9 \
{\rm cm \ s^{-1}}$, but still fast enough at $0.9 \times 10^9 \ {\rm cm \
s^{-1}}$ to go around the star prior to the transit through the star of the
inwardly propagating weak shock.  Our simulations resolve the subsonic region
behind the reaction front in the detonation wave.  The 2D nucleosynthesis
is shown to be consistent with a truncated 1D Zeldovich-von Neumann-D\"oring
(ZND) calculation at the slower detonation speed.  The ashes from the lateral
detonation are typically He rich, and consist of predominantly $^{44}$Ti,
$^{48}$Cr, along with a small amount of $^{52}$Fe, with very little $^{56}$Ni
and with significant $^{40}$Ca in carbon-enriched layers.  If this helium
detonation results in a Type Ia Supernova, its spectral signatures
would appear for the first few days after explosion.

\end{abstract}


%%%%%%%%%%%%%%%%%%%%%%%%%%%%%%%%%%%%%%%%%%%%%%%%%%%%%%%%%%%%%%%%%%%%%%%%
\section{Introduction}

Recent theoretical insights \citep{bildsten07, shenetal10,
waldman10,woosleykas11} and observations of very rapidly evolving
thermonuclear supernovae such as 2002bj \citep{poznanski10}, SN2010X
\citep{kasliwal10}, SN 1939B, and SN 1885A \citep{perets11} motivate an
investigation of the outcome of unstable burning of intermediate mass
($0.01M_\odot<M_{\rm He}<0.2M_\odot$) Helium shells on accreting white dwarfs
(WDs). \citet{shenbild09} showed that the burning initially
proceeds as a spherically symmetric slowly evolving (minutes to hours)
convective layer [see also \citet{woosleykas11}]. For even  lower shell
masses (such as occur in stars on the Asymptotic Giant Branch), the
convective burning causes radial expansion adequate to increase the 
evolutionary timescales to times long enough for confidence in a
one-dimensional hydrostatic calculation. However, there is a critical helium
shell mass ($M_{\rm He}>0.1 (0.01) M_\odot$ for a WD mass of
$M=0.6(1.1)M_\odot$) above which the convective burning timescales become so
short that a dynamical mode of burning will occur \citep{shenbild09}. 

As highlighted by many \citep{bildsten07,shenetal10,woosleykas11}, the
dynamically burning He shell differs from the well studied case of dynamic
carbon burning in WD cores during Type Ia SNe, as the finite gravitational
acceleration, $g$,  in the He shell rapidly forces mixing of any unstable
density gradients from a radially propagating deflagration
\citep{TimmesNiemeyer00}.  This prohibits
the localization of a well defined and propagating deflagration front.  If
detonation is the result, we expect that a lack of synchronicity across the
star will lead to laterally propagating detonation wave(s)  from the ignition
site(s).  \citet{fink07,fink10,simetal12} revived the exciting possibility
\citep{livglas91,woosweav94,livarnett95} that the weak shock that penetrates
the underlying WD from a laterally propagating He detonation  will
strengthen enough (even for a $M_{\rm He}=0.01M_\odot$ shell) 
at the off-center focus point to trigger a C/O detonation in the WD. 
They have argued that nearly all He detonations will lead to a complete
explosion of sub-Chandrasekhar C/O mass WDs. This motivated theoretical
modeling of such explosive events \citep{simetal10,kromer10,
waldman10,woosleykas11, simetal12} finding similarities to the
width-luminosity relation of Type Ia SNe and potentially  offering new
insights on unusual events [e.g. SN 2005E \citep{perets10}, PTF09dav
\citep{sullivan11}, and others \citep{kasliwal11}]

Though uncertainties remain \citep{shenetal10,woosleykas11} as to the exact
criteria  that must be met to initiate a detonation in these lower density
($\rho<10^6 \ {\rm g  \ cm^{-3}}$) He shells, we work here to resolve how the
lateral propagation will be impacted by the radial expansion that occurs
during the burning.  The radial blow-out timescale is $H/c_s\sim 0.01 \ {\rm
s}$ for a shell of thickness $H\approx 10^7 \ {\rm cm}$, where the sound
speed, $c_s\approx 10^9 \ {\rm cm \ s^{-1}}$, is set by the post-shock
temperature of the detonation front. This timescale decreases for thinner
layers, whereas the nuclear burning timescale in the detonation reaction zone
increases.  If a propagating detonation wave remains stable under incomplete
burning, it would reduce the propagation speed to values lower than the
Chapman-Jouguet (CJ) speed of $1.5\times 10^9\ {\rm cm \ s^{-1}}$ calculated
for a complete burn of He to $^{56}$Ni \citep{simetal12}.  At even lower
densities (or shell thicknesses), the blow-out occurs so rapidly that the
propagation could be quenched.

We begin in \S\ref{sec:blowout} by showing a 2D simulation of
a steadily propagating detonation in a thin helium layer for an example case
motivated by the AM~CVn accretion scenario.  In order to investigate the
robustness of our numerical techniques and better understand the incomplete
burning that occurs, we undertake two sets of comparisons.  In
\S\ref{sec:transientdetonation}, we perform 1D and 2D simulations, with
enough resolution to resolve all the burning scales, investigating transient
strengthening detonations in uniform pure He.  We find that the incompletely
burned strengthening detonation provides a useful comparison to the
blowout case.  In \S\ref{sec:zndcomparison}, we compare our simulations to
integrations that give the structure of 1D steady-state detonations
(ZND\footnote{Throughout we use Zeldovich-von Neumann-D\"oring (ZND) to refer to
the eigenvalue method for computing the spatial structure of steady state
detonations.
See, e.g., \citet{Khokhlov89} for a convenient form of the equations
integrated and \citet{FickettDavis79} for a broader discussion.})
performed with a large (200-nuclide) reaction network and find qualitative
agreement.  Finally, in \S\ref{sec:conclusions}, we discuss astrophysical
implications and future work.

%%%%%%%%%%%%%%%%%%%%%%%%%%%%%%%%%%%%%%%%%%%%%%%%%%%%%%%%%%%%%%%%%%%5
\section{Surface Blowout Induces Freeze-out of Intermediate Products}
\label{sec:blowout}

Our fiducial model for the multidimensional runs was motivated by a time
dependent $\dot M$ calculation with
$\MESA$\footnote{http://mesa.sourceforge.net} \citep{paxton11} that had a
declining $\dot M$ as expected from a degenerate He donor in an AM CVn binary
\citep{Bildetal06,bildsten07}. This exploratory calculation had $10$ He flashes
at very high accretion rates on the $M=1.0M_\odot$ WD, with about two-thirds
of the accreted material leaving during the He novae. The remaining
carbon-oxygen ashes from the He burning creates a hot ($T\approx 8\times 10^7
\ {\rm K}$) layer of $\approx 0.015M_\odot$. The final flash occurred when $\dot
M\approx 6\times 10^{-8}M_\odot \ {\rm yr^{-1}}$, with a freshly accumulated
helium layer mass of $0.028M_\odot$. At this $\dot M$, the temperature
maximum occurs above the base of the He shell, placing the base of the
convective zone at a mass coordinate inward from the surface of
$0.014M_\odot$. Our fiducial model is comparable to two $M=1.0M_\odot$ WD
models of \citet{woosleykas11}. The
first (10AA) had a cold C/O core ($T<3\times 10^7 \ {\rm K}$) and $\dot
M=7\times 10^{-8}M_\odot \ {\rm yr^{-1}}$, giving an accumulated mass of
$M_a=0.052M_\odot$ and a convective layer of $0.0262M_\odot$, thicker than
our He layer due to the colder core. The second (10HB) had a hot C/O core
($T\approx (7-8)\times 10^7 \ {\rm K}$) and $\dot M = 5\times
10^{-8}M_\odot \ {\rm yr^{-1}}$, giving an accumulated mass of
$M_a=0.022M_\odot$ and a convective layer of $0.022M_\odot$, much closer to
our fiducial.  Compared to the models shown in their Figure 2, our model has
a peak luminosity of $10^{46.6}$~erg~s$^{-1}$ when the conditions at the base
of the convection zone are $\rho_6=0.14$ and $T_8=5.0$.  This places our
model above their threshold for shell detonations.

\begin{figure}
\plotone{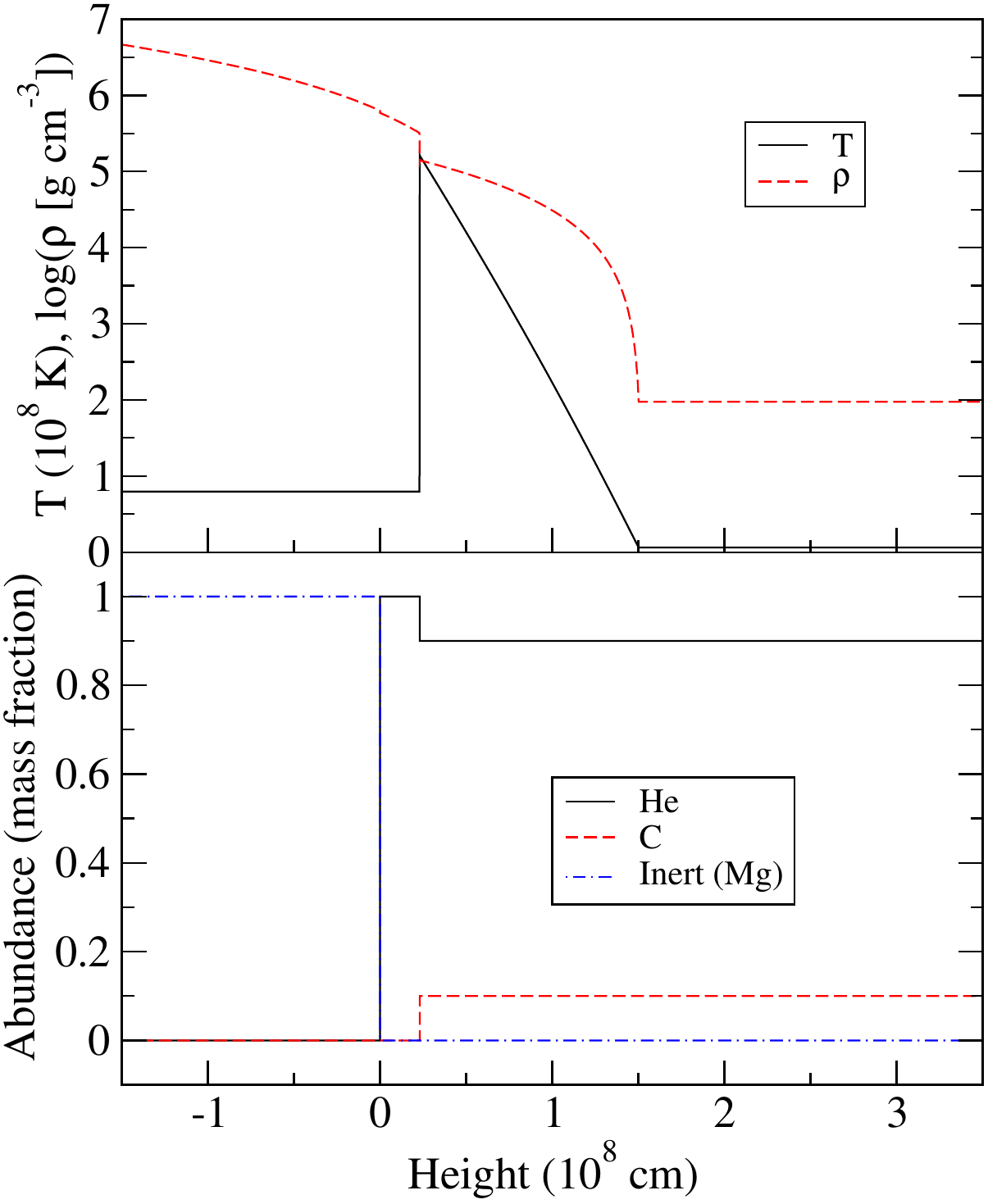}
\caption{\label{fig:init_vert_profile}
Initial vertical profile on the domain of temperature, density and abundances.
}
\end{figure}

For our simulation, we used the temperature and density structure
when the timescale for temperature change in the convective
burning layer was at its minimum, about 5 seconds.
This calculation is intended to demonstrate one case in which the
burning scales are long enough for blowout to truncate the reactions. 
This radial (vertical)
structure is shown in Figure \ref{fig:init_vert_profile}.
% can be seen at the left edge, ahead of the detonation front,
%in Figure \ref{fig:blowout}, which shows a snapshot of the detonation
%propagation 0.5 seconds after the detonation was ignited at
%$x=4.5\times 10^8$ cm.
 The initial helium layer is in hydrostatic
balance, with the hot ($T_9=0.52$), carbon-enriched ($X_{12}=0.1$ due
to He burning) convection zone above the colder ($T_9=0.08$) pure He
layer. The base of the cold He layer defines the zero of the vertical
coordinate, and has a density of $\rho_5=5.89$. The cold layer extends
up to $\approx 0.25\times10^8$~cm, and has a column depth of
$10^{13}$~g~cm$^{-2}$ or $\approx0.01M_\odot$, while the convective
layer extends to $\approx 1.4\times 10^8$~cm and has a similar mass. Above
this is a low-density "fluff" layer outside the star extending to the edge of
the grid.  We work
in plane parallel, with $g=7.32\times10^8~\rm cm~s^{-2}$,
appropriate to the $1M_\odot$ WD of radius $R=4.3\times10^8$~cm. The helium
is sitting on a presumed (for this simulation) inert core.

Reactive hydrodynamics simulations were performed using
FLASH\footnote{http://flash.uchicago.edu} \citep{Fryxell+00} and
utilized the ``approx13'' nuclear reaction network
\citep{TimmesHoffmanWoosley00} that accounts for the fast
$(\alpha,p)(p,\gamma)$ side channels relevant to alpha burning
\citep{Timmes+00}.  We used the piecewise parabolic method (PPM;
\citealt{ColellaWoodward84}) for compressible hydrodynamics.  Reactions were
disabled in regions of the grid within a detected shock
\citep{FryxMuelArne89}.  The hydrodynamic and reaction evolution are operator
split and the temperature is held fixed during the reactions.
A hydrostatic lower boundary condition is used
\citep{Zingale+02}, implemented by Townsley and included in FLASH
3.3. The plane-parallel calculation is performed with a uniform
$1.22\times 10^5$cm grid of cells with large lateral width of $5\times
10^8$cm. We do not
initialize any convective motion, rather we just use the thermal
profile implied by a constant entropy and abundance convective zone.

\begin{figure*}
\epsscale{1.15}
\plotone{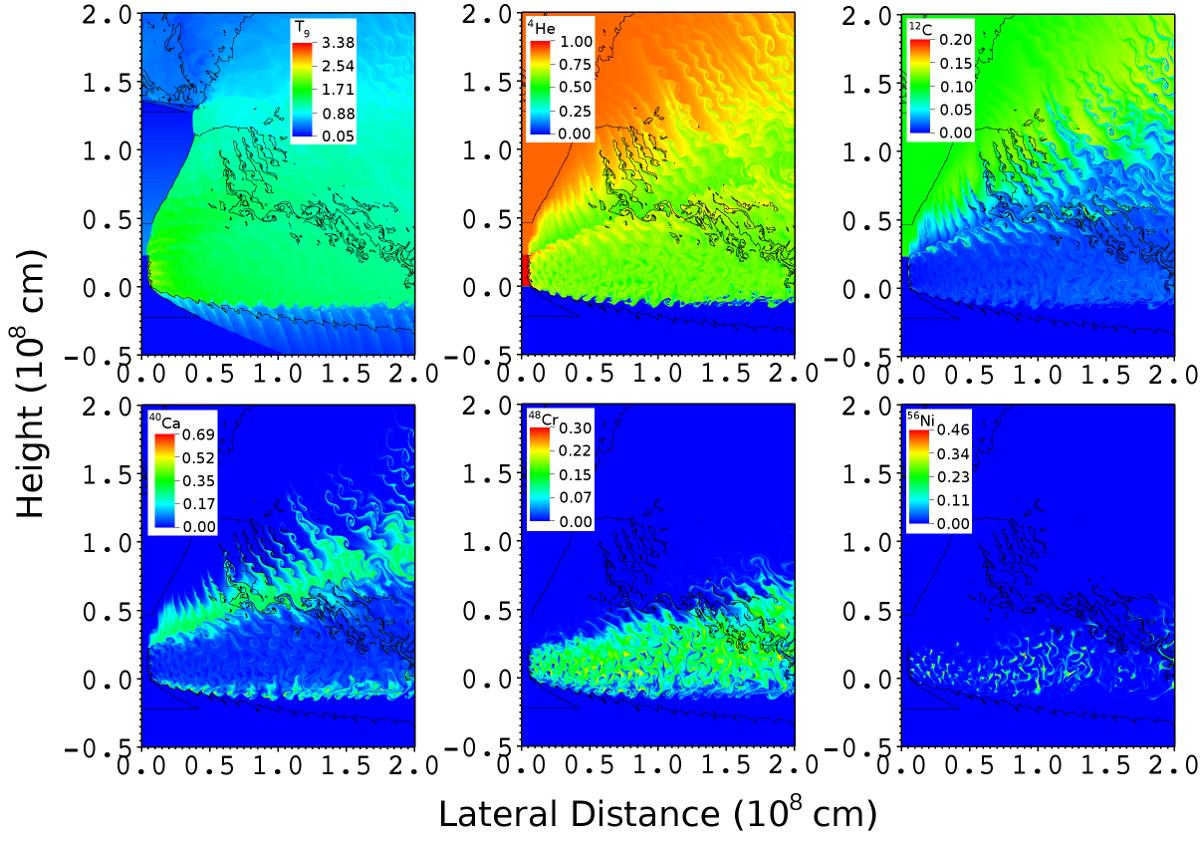}
\caption{\label{fig:blowout}Steady-state temperature and abundance (in mass
fractions) structure of a surface detonation propagating from right to left
in the accreted helium layer of a WD.  Prior to ignition, a hot, actively
burning convection zone (0.25-1.4 on the vertical (radial) axis) enriched in
$^{12}$C ashes sits above a cold helium layer (0.0-0.25) atop an inert core.
Density contours are shown at $10^3$, $10^4$, $10^5$ and $10^6$~g~cm$^{-3}$.
This snapshot is from 0.5~s after the detonation was ignited at $x=4.5\times
10^8$ cm.  Only a portion of the computational domain near the detonation
front is shown.
}
\end{figure*}

We numerically initiate the detonation by raising the temperature in the
convective zone to $T_9=3$ in a circular region of radius $10^7$~cm, centered
$1.2\times10^7$~cm above the base of the convective layer. This does not
directly ignite a detonation, but rather creates a shock wave that traverses
the cold layer; reflects from the inert core, and emerges out as a detonation
wave.\footnote{This brings up a shortcoming of this calculation, which is
that this interaction led to immediate ignition of C/O when we allowed for
carbon fusion. This is an artifact of our initiation scheme, and led
us to use $^{24}$Mg for the underlying layer in this initial study.
%Maybe we turn off Carbon fusion underneath??
}
The resulting detonation front is shown
in Figure \ref{fig:blowout}, which shows a snapshot of the detonation
propagation 0.5 seconds after the detonation was ignited at
$x=4.5\times 10^8$ cm.
  The large lateral width of our simulation allows for the
detonation wave to propagate far from the initiation site,
demonstrating a steady-state form of much smaller spatial dimensions
($\lesssim 10^7$cm). The resulting speed of the detonation front is shown in
Figure~\ref{fig:det_speed_plot} by the points labelled ``Blowout" and reaches
8.7$\times 10^8$~cm~s$^{-1}$ for several tenths of a second as it crosses our
large domain. 

The resulting composition profiles are shown in
Figure~\ref{fig:blowout}.  The burning is far from complete, with the
$^4$He abundance far behind the detonation front only falling to
$\approx 0.5$. Scarcely any $^{56}$Ni is produced, while more
significant amounts of slightly lower $Z$ elements are made, including
$^{48}$Cr (bottom middle panel) and $^{44}$Ti, but relatively little
$^{52}$Fe (not shown). The nucleosynthesis in the hot, carbon-enriched
layer gives a similar amount of unburned helium, producing generally
lower $Z$ products such as $^{40}$Ca (lower left panel), and appears
to be a detonation wave partially supported by the overpressure from
the underlying detonation. We performed a similar calculation in
which the overlying layer had no carbon, and the detonation speed was
slightly lower, 8.5$\times 10^8$~cm~s$^{-1}$, but all other features
were qualitatively similar.
The detonation front (vertically extending to $0.5\times 10^8$ cm) is
cellular, with the strongest shock-fronts propagating at an angle with
respect to the overall leftward velocity. These cross-propagating
shocks traverse both the cold pure helium and the carbon enriched
layer above.
%Based on comparison to a lower resolution 2-dimensional
%run, the cellular structures in this blowout calculation may still be
%larger than they should be. This does not appear to adversely affect
%the nucleosynthetic outcome, as the important burning stages active
%during the truncation are resolved.

We hypothesize that the surface blowout has induced a eigenvalue-type
(sometimes called ``pathological") detonation solution
\citep{FickettDavis79} in which the sonic point, where the flow
becomes supersonic with respect to the shock front, is inside the
reaction zone. This truncates the burning while still preserving a
self-propagating detonation structure between the shock front and the
sonic point (actually sonic locus in multiple dimensions).  This is
surprising, as discussed below in
Section~\ref{sec:transientdetonation}, a detonation in uniform pure
helium at the density at the base of the helium layer 
has a ZND length (to reach pure $^{56}$Ni) much longer than our grid.

%%%%%%%%%%%%%%%%%%%%%%%%%%%%%%%%%%%%%%%%%%%%%%%%%%%%%%%%%%%%%%%%%%
\section{Large and Transient Detonation Structures}
\label{sec:transientdetonation}

As we have shown (see also Sim et al. 2012), the low helium densities
near the WD surface allow matter to blowout on a timescale comparable
to the burning time. This is in contrast with earlier work done on
Helium detonations in denser environments. For example,
\citet{Khokhlov89} performed ZND
calculations for pure helium at $\rho_5=50$, finding complete burning
by the time (and distance, $10^8$~cm) the flow behind the detonation
front reaches the sonic point and a speed, $1.5\times
10^9$~cm~s$^{-1}$, consistent with the Chapman-Jouget (CJ) value for
complete burning. However, at lower densities, the detonation
structures become much larger than the vertical (and even horizontal)
scale of the outer layers of a WD \citep{TimmesNiemeyer00}, requiring a new approach. It also
raises the possibility that a propagating detonation is not even
possible in these thin helium layers, a quenching outcome well known
in chemical detonations.

We begin by investigating how marginally ignited detonations
strengthen towards the steady-state CJ solution and how this depends
on composition. At low densities, these solutions will extend to
physical dimensions much larger than can be accommodated on a WD, but
will enable comparisons to ZND calculations that help us understand
the likely effects of a finite geometry.  In order to facilitate
direct comparison with ZND calculations, we work in a 1-dimensional
plane parallel geometry in a medium of uniform density, temperature
(typically at $T_9=0.1$) and composition, sometimes including a small
carbon fraction. The detonation is ignited by setting the temperature
close to the boundary at $T_9=3$ with a discontinuous fall to the
ambient temperature. We usually use the smallest hot region that leads
to a propagating detonation, decrementing a factor of 10 at a time.
We perform these simulations at two
different resolutions, a coarse resolution to allow the detonation to
reach steady state, and a fine resolution which resolves all the
burning stages.  Adaptive mesh refinement is used in all cases, so
that the resolution is actually the minimum cell size.

\begin{figure}
%\epsscale{0.8}
\epsscale{1.1}
\plotone{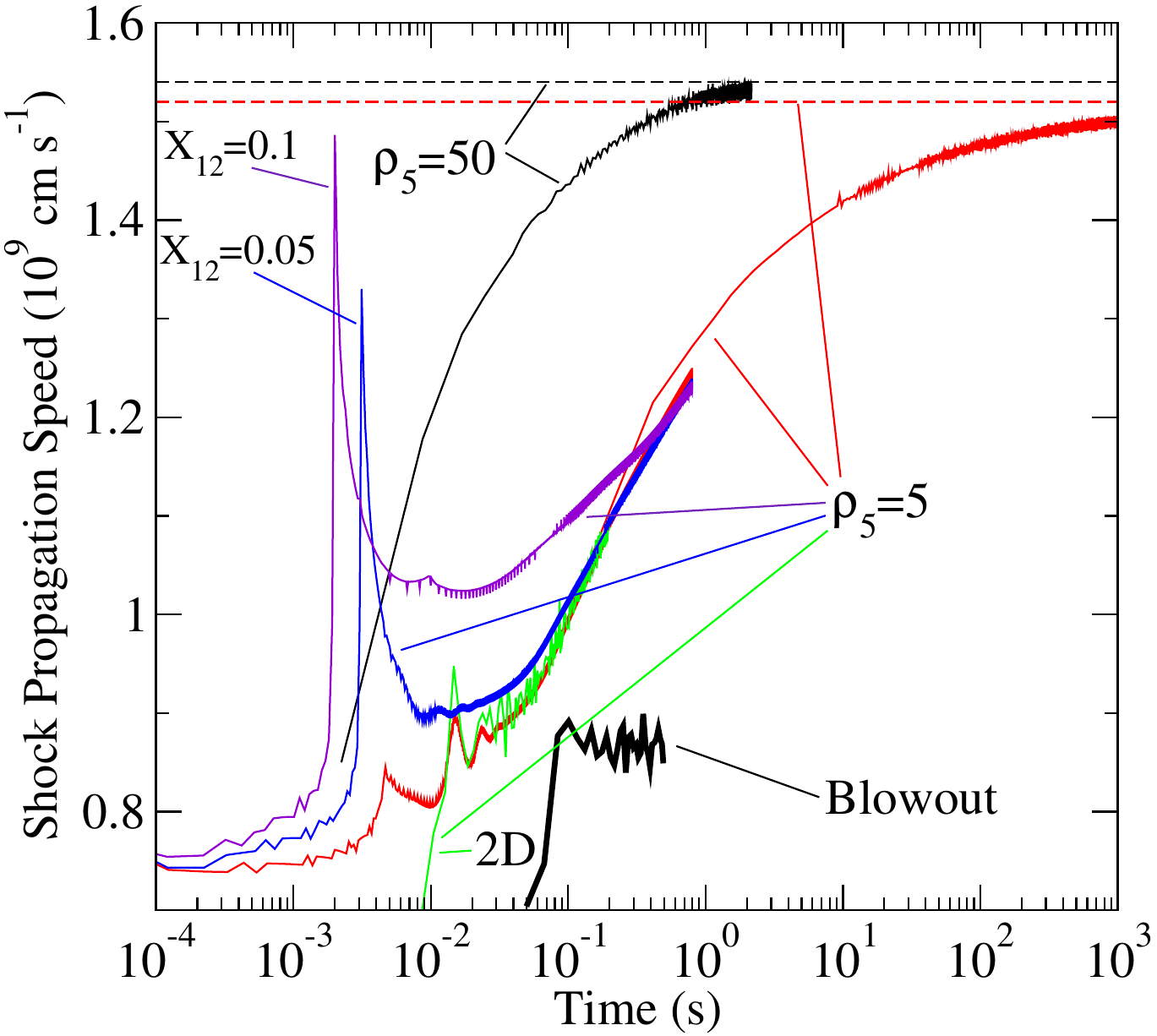}
\caption{\label{fig:det_speed_plot} Detonation shock velocity as a function of
time for detonations in helium.  Two densities for uniform fuel are shown,
$5\times 10^6$~g~cm$^{-3}$ (solid black) and $5\times 10^5$~g~cm$^{-3}$ (1D: solid red,
2D: green), and dashed lines show the corresponding steady state CJ speeds.
The speed for material containing some $^{12}$C in fuel, $X_{12}=0.05$ (blue)
and $X_{12}=0.1$ (purple) is initially higher but asymptotes to similar
behavior.  The propagation speed of the detonation wave moving through the
helium surface layer on a WD (thick black) stabilizes well below the CJ
speed.
%  A White
%Dwarf is only large enough for
%the detonation to propagate for roughly 1 second, therefore Helium
%detonations at densities $\lesssim 5\times 10^6$~g~cm$^{-3}$ will not have
%strengthened to the CJ form.
%  These show a higher speed, and therefore
%likely more energy release, from the structure formed by 0.1~s, but
%eventually also asymptote toward the CJ-type solution.
%  These show a fairly stable
%speed significantly lower than the CJ speed, consistent with the truncation
%of the burning from blowout precluding the full strengthening of the
%detonation.
}
\end{figure}

We start by comparing to Khokhlov (\citeyear{Khokhlov88},
\citeyear{Khokhlov89})'s calculations at $\rho_5=50$ and $T_9=0.2$ for
the initial state. The speed at which the shock front propagates as
the $10^5$~cm resolution simulation progresses is shown by the black curve in Figure
\ref{fig:det_speed_plot}. This detonation is of the CJ-type and the CJ
speed (as determined by a calculation containing only the 13
$\alpha$-chain elements) is shown by the horizontal dashed line at a
speed of $1.54\times 10^9$~cm~s$^{-1}$. This is also the speed
obtained by interpolating in Table IV in \citet{Khokhlov88}, and we
confirm his spatial profiles. Though it only takes $0.01 {\rm s}$ to
reach $80\%$ of the CJ speed, the full speed is not attained for about
one second.

We now consider a case with $\rho_5=5$, similar to the surface
detonation calculation presented above. The evolution of the shock
velocity (solid red) is shown in Figure \ref{fig:det_speed_plot}, and
is qualitatively similar to the higher density result. Note that there
are two red curves, one at a coarse ($4.8\times 10^6$ cm) resolution
and one much finer resolution (1000~cm). We performed the fine
resolution simulations at a variety of ignition sizes, finding that
ignition sizes of $10^5$~cm and below fail to produce a propagating
detonation. We therefore use an ignition size of $10^6$~cm for all the
fine simulations shown here. Similar to the higher density result,
this calculation asymptotes to the CJ speed, but only after 1000
seconds and a propagation length $\approx 10^{12}$~cm, larger than the
whole WD.

We also explored the effects of a modest fraction of $^{12}$C in the
fuel, which \citet{shenbild09} suggested may provide a faster
reaction channel via $^{12}$C($\alpha$,$\gamma$)$^{16}$O.  We
performed simulations with $X_{12}=0.05$ and $0.1$, shown by the blue
and magenta curves in Figure~\ref{fig:det_speed_plot}.
Before about $10^{-3}$~s, the shock velocity is that of the leading shock from
our ignition, not yet a detonation.  The region between the contact
discontinuity at the edge of our ignition region and this leading initiation
shock is effectivly preconditioned by the shock propagation, and runs away to
ignite the detonation.  The spike in propagation speed at a few$\times
10^{-3}$~s is when this newly-formed detonation first passes the leading
shock. It leaves this pre-heated region with an enhanced speed due to the
preheating, falling quickly as the detonation propagates into now unheated
fuel.
Due to the presence of carbon, there is
a continued enhancement of the propagation speed up to about 0.1~s, potentially
relevant if carbon is present in the accumulated cold He layer. As the 
detonation continues to strengthen, the speed becomes dominated
by the consumption of $^4$He, and asymptotes toward the CJ solution.

To investigate the cellular structure of these detonation fronts, we
also performed a 2-dimensional calculation. The configuration was
plane-parallel in a uniform medium, but now with random 1\% density
perturbations near the ignition front to seed the cellular
instability. The domain width transverse to the detonation front was
$10^7$ cm with periodic boundary conditions, adequate to accommodate
several detonation cell widths, though the cell width appeared to grow
with time as the detonation strengthened.  The resolution for the 2-d
simulation was about 5000~cm, enough to marginally resolve the fastest
stage, helium capture on produced carbon. Even so, the evolution of
the detonation speed (green curve in Figure~\ref{fig:det_speed_plot})
agrees quite well with the 1-d simulations, giving us confidence that
our 1-d simulations are robust enough to compare to the inherently
multi-dimensional blowout case.

This 2-dimensional uniform density simulation also allows us to study
resolution effects in a more controlled context.  By comparing a
calculation at $7.8\times 10^4$ cm resolution to the one at 5000~cm,
it appears that size of the cellular structure is influenced by
resolution, with larger distance between cross-propagating shock
fronts at coarser resolutions.  However this does not have a
significant impact on the resolved nucleosynthetic structures 
such as for the higher-$Z$ elements (i.e. Ca, Ti, Cr, Fe and Ni). 

We find that weakly ignited (initially underdriven) detonations in
helium strengthen toward the steady-state solution over time, but if
the overall detonation structure -- that is the thermodynamic and
compositional profiles in space -- is large, it will take a
significant run of space for the detonation to take on this steady
state structure.  The incomplete burning structure created by the
blowout near the WD surface is similar to the early phases of these
strengthening detonations, providing a valuable comparison.

%%%%%%%%%%%%%%%%%%%%%%%%%%%%%%%%%%%%%%%%%%%%%%%%%%%%%%%%%%%%%%%%%%
\section{Verification between ZND Calculations and FLASH Simulations}
%\section{Scales for Intermediate Products (in strengthening He detonations)}
\label{sec:zndcomparison}

We have established that the helium detonation blowout at low
densities reduces the available burning time, inhibiting the complete
burning of helium and highlighting the importance of accurately
calculating the production of intermediate mass products.  We begin by
exploring how the details of the reaction network and calculation
method impact the nucleosynthesis. So as to save computational time,
the FLASH calculation uses the ``approx13''
\citep{TimmesHoffmanWoosley00} reaction network, a simplified 13
nuclide  network with steady-state approximations of heavier $p$-rich
nuclei along $(\alpha,p)(p,\gamma)$ channels and their inverses. We
check whether this approximation leads to any substantial
nucleosynthetic differences by comparing to 1D ZND calculations
performed with a 200 nuclide nuclear network.

In a standard ZND computation, the eigenvalue detonation speed (or,
equivalently, shock strength) is obtained when the position in the flow where
the heat release function \citep[see e.g.][]{Khokhlov89} is zero coincides
with the flow velocity being sonic with respect to the shock front.  This
gives a physically consistent transition from the detonation structure to the
following flow beyond the sonic locus.  However, we cannot perform a standard
ZND computation for anything but the true steady state, which in this case is
a far stronger detonation than either the early transient or blowout cases.
Instead, we perform the first portion of a ZND integration, starting from the
shock, for the known shock strength drawn from simulations and truncate our
integration near the sonic point where it would become singular.

\begin{figure}
\epsscale{1.1}
\plotone{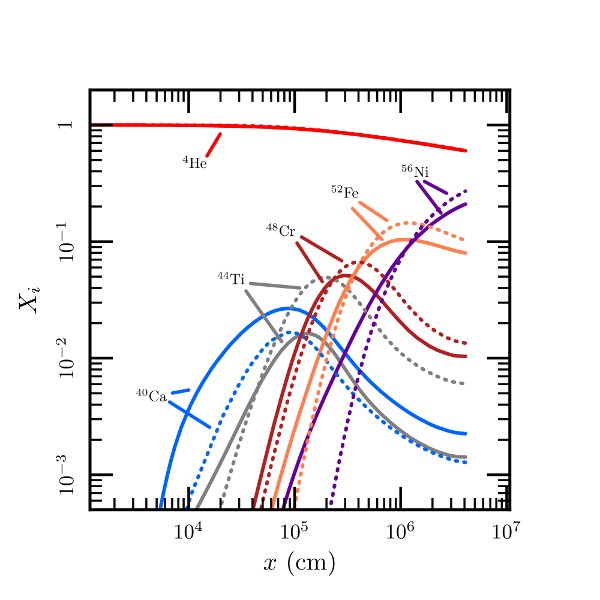}
\caption{
Comparison of post-shock nucleosynthesis of detonations in pure He computed in the ZND formalism using two different reaction networks. The dotted lines are calculated using the aprox13 reaction network and the solid lines from the 200-isotope network. The detonation velocity was $D=1.09\times 10^9$ cm/s for each run and the ambient conditions were $\rho_0 = 5\times 10^5\ {\rm g\ cm^{-3}}$ and $T_0=10^8\ {\rm K}$. \label{fig:znd_net_comp}}
\end{figure}

The first step is to examine the nucleosynthetic differences between a
ZND calculation with approx13 and the 200 nuclide nuclear network at
the same conditions ($\rho_5=5$ and $T_8=1$).  We choose to initialize
the comparison ZND integrations with the detonation velocity achieved
at 0.2 seconds in the high resolution uniform medium 2D FLASH run,
$1.09\times 10^9$~cm~s$^{-1}$.  This velocity is still lower
than the steady-state CJ detonation velocity, therefore the ZND
integration will terminate when the downstream velocity in the shock
frame is equal to the sound speed, here at a distance of $4\times
10^6$~cm. The resulting nuclide abundance distributions are shown in Figure
\ref{fig:znd_net_comp}, showing that approx13 provides a suitable
reproduction of the results obtained from the 200 nuclide network.
$^{44}$Ti is overproduced by about a factor of two at intermediate length
scales, indicating that this nuclide is likely slightly overpredicted in our
yields.

\begin{figure}
\epsscale{1.1}
\plotone{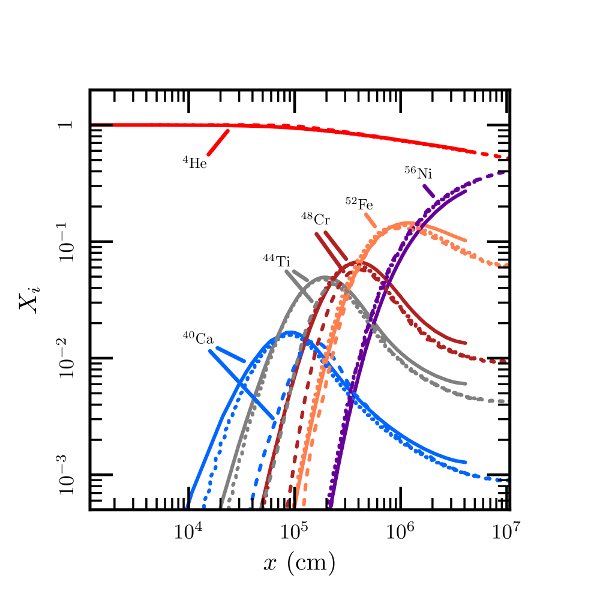}
\caption{
Comparison of laterally averaged 2-dimensional (dashed) and 1-dimensional
(dotted) FLASH simulations of the He detonation structure after $0.2$ s of propagation. Also shown is a ZND calculation (solid), which uses the FLASH detonation speed at $0.2$ s, $1.09\times 10^9$~cm~s$^{-1}$. Ambient conditions were $\rho_0 = 5\times 10^5\ {\rm g\ cm^{-3}}$ and $T_0=10^8\ {\rm K}$. Both FLASH and ZND are using the approx13 reaction network. \label{fig:znd_flash_2d}}
\end{figure}

A next check is to verify that the intermediate element production is
consistent between the uniform density simulations and the ZND
integration, both using the approx13 network.
We compare, in
Figure~\ref{fig:znd_flash_2d},
the simulation (dotted line for 1D), which captures the
beginning phase of the detonation while the shock is still
strengthening, to the truncated ZND integration (solid line for ZND).
The excellent comparison shows that the truncated ZND is a
suitable approach to understanding the strengthening detonations.
Additionally, we need to relate the 2D simulations to those in 1D.
The laterally averaged 2D results are shown by the dashed line in
Figure \ref{fig:znd_flash_2d}, remarkably similar to the 1D results.
Apparently the cellular nature of the 2D detonation did not
significantly change the burning length scale from the 1D case.

We have shown consistency in the detonation structure computed by all
of the following: (1) a truncated ZND calculation with a large
network, (2) a truncated ZND calculation with the aprox13 network, (3)
1D hydrodynamic simulations of the transient phase of a strengthening
detonation that resolve all relevant reaction length scales, (4) 2D
hydrodynamic simulations that nearly resolve all relevant length
scales. These consistencies give us confidence in the propagating
solution revealed by the blowout simulations, but we have yet to use
these truncated ZND calculations to fully diagnose (and maybe predict)
the blowout results (both nucleosynthetic yields and speeds).

%%%%%%%%%%%%%%%%%%%%%%%%%%%%%%%%%%%%%%%%%%%%%%%%%%%%%%%%%%%%%%%%%%%%%%
\section{Conclusions}
\label{sec:conclusions}

We have shown that a thin helium layer on the surface of a WD can host
a steady-state, laterally propagating detonation at low densities
($<10^6$~g~cm$^{-3}$). Under these conditions, the detonation takes on
the character of a eigenvalue-type detonation in which the sonic locus
occurs well before complete helium burning is achieved.
The stability and compact structure of the resulting detonation is
surprising because the size of the reaction zone of a steady-state
detonation, which burns to completion,
at these densities ($\gg10^8$~cm) is larger than the layer thickness ($\sim
10^7$~cm) or even the star itself.
Our 2D
reactive hydrodynamic simulations resolved the subsonic region behind
the shock, exhibited the radial blowout of the hot material, and
primarily yielded intermediate mass products (i.e. $^{44}$Ti and
$^{48}$Cr) and relatively little $^{56}$Ni. The incomplete burning had
a spatial scale much smaller than the vertical thickness of the helium
layer, and a detonation speed of $0.9 \times 10^9$~cm~s$^{-1}$,
lower than the Chapman-Jouget speed of complete burning, $1.5\times
10^9$~cm~s$^{-1}$.

This lower detonation speed reduces the strength of the inwardly
moving shock that may ignite a detonation in the carbon-oxygen core
and cause a Type Ia SNe \citep{fink07,fink10,simetal12}. The slower
speed moves the convergence point for these weak shocks further
towards the surface, potentially making it harder for the carbon to
detonate.  If a carbon detonation still occurs, the reduction in high-$Z$
elements that we found for the helium layer would  allow for colors and
spectral features consistent with normal Type Ia SNe over most of their
evolution \citep{kromer10,woosleykas11,simetal12} for even larger helium
layer masses.  Modest carbon enrichment in the helium layer was found here to
produce significant amounts of $^{40}$Ca, which, if the underlying core
detonated, would appear in spectra at high velocities with several possible
sources of asymmetry, a common feature of SNIa
\citep{kasenetal03,mazzalietal05,tanakaetal08,foleyetal12}.  It is
certainly the case that the extremely low-mass helium shell considered here
would only effect the earliest time colors \citep{woosleykas11,piro12},
highlighting the importance of obtaining early observations of SNe~Ia
\citep{nugentetal11,foleyetal12}.

If the carbon does not detonate, then the only observable signatures
would be from the ejected ashes of the helium detonation
\citep{bildsten07,shenetal10}, a ``.Ia'' supernovae.  For this
$M=M_\odot$ WD, the gravitational binding energy of the material at
the surface is $3\times 10^{17} {\rm ergs \ g^{-1}}$, and rough
integrals over the 2D simulation yielded burning products of
$X_4=0.66$, $X_{12}\approx 0.054$, $X_{40}\approx 0.066$,
$X_{44}=0.079$, $X_{48}\approx 0.053$, $X_{52}=0.026$,
$X_{56}=0.0067$, and a smattering of other elements. This incomplete burning
releases $\approx 4.0 \times 10^{17} \ {\rm erg \ g^{-1}}$, so that the ashes
are unbound and reach infinity with an average velocity of $4500\ {\rm km \
s^{-1}}$, a rather low velocity compared to normal supernovae. The small
ejecta mass ($0.028M_\odot$) will
lead to a short duration (likely less than 2 days near peak light) and
the small amount of long-lived radioactivities imply a low luminosity
($\ll 10^{42} \ {\rm erg \ s^{-1}}$).  This rather low energy release brings
up the interesting possibility that, for thin shells, a detonation might not
unbind the surface layers, though the CO interior might still shock-ignite.

There remain several important uncertainties.  The greatest is related
to the initiation of the detonation in the convective helium burning
layer on the WD surface \citep{shenetal10,woosleykas11}. We assumed
here that a detonation was ignited and just considered its propagation
and stability. We intend to first explore the dependence of the
detonation speed on both the He layer thickness and density. Coupled
with our truncated ZND calculations (see section~\ref{sec:zndcomparison}), we
hope that these insights will
yield a direct approach to predicting the minimum possible layer that
will allow for a propagating detonation. This more complete
understanding of the mechanism by which the blowout of the surface
layer leads to an eigenvalue-type stable detonation solution should
allow for predictions of nucleosynthetic outcomes across a broad range
of parameter space without expensive multidimensional simulations.

%Typically in a terrestrial explosive,
%such a vigorous expansion would quench the detonation.
%(TODO references?, so we can be more specific. mesh with intro/conclusion?)
%We believe that
%the helium case is different because
%many of the early reaction steps occur quite quickly despite the large size
%(slow rate to reach complete burning) of the equilibrium detonation solution
%in a uniform medium.

%At this early stage in our theoretical understanding of the transition from a
%hydrostatic calculation to the onset of a more dynamical outcome, we have decided to address one distinct issue that is calculable. 

%This could quench the detonation front, analogous (but not completely so) to the quenching of chemical detonations under strong curvature. Hence, we address here the distinct question o

%Here is a good place to talk about the weaker shock going into the C/O.
%these results are for pure He and mention that small amounts of carbon will
%change things.

%%%%%%%%%%%%%%%%%%%%%%%%%%%%%%%%%%%%%%%%%%%%%%%%%%%%%%%%%%%%%%%%%%%%

\acknowledgements

We thank Ryan Foley for helpful discussions and Bill Paxton for MESA
calculations.
Some of the software used in this work was in part developed by the
DOE-supported ASC/Alliances Center for Astrophysical Thermonuclear Flashes at
the University of Chicago.  We thank Nathan Hearn for having made his
QuickFlash analysis tools publicly available at
http://quickflash.sourceforge.net.  We used Frank Timmes'
nuclear reaction network (http://cococubed.asu.edu) for preliminary
calculations.  Some simulations presented in this work were run
on the Ranger supercomputer at the Texas Advanced Computing Center as part of
the Extreme Science and Engineering Discovery Environment (XSEDE, formally
TeraGrid), which is supported by National Science Foundation grant number
OCI-1053575.  This research has been supported by the National Science
Foundation under grants PHY 11-25915 and AST 11-09174. 

%Dean's What's next:
%\begin{itemize}
%\item Show that what we say works out: the solution we are seeing is a sort
%of eigenvalue solution where the analog of dissipation is provided by the
%blowout of material.  This can be done by showing that ZND equations with a
%blowout term in them matches the flash hydro (use particles for this).
%\item Do some real nucleosynthesis and apply to some astrophysical situation.
%i.e. post-process and calculate some ejecta profiles.
%\item Parameter space survey: how do things vary with density, carbon
%abundance, WD mass (surface gravity) etc.
%\item Do a good calculation of inward shock strengths -- apply to
%double-detonation carbon ignition.  I can do the supernova calculation too.
%
%\item The elephant in the room: the ignition process.  HARD.  Makes some of
%the above harder too.  I am probably going to continue to propose to get some
%money to do this as it requires some work with convection zone computations
%and such.
%\end{itemize}
%
%Kevin's What's next:
%\begin{itemize}
%\item Perhaps more ZND runs with carbon (what conditions allow the CJ detonation to fit in the star?)
%
%\item Debug ZND code in MESA (why is it so slow?) and make comparisons to networks computable in Timmes' code.
%
%\item Finalize prescription for treating blowout in 1D and agree on new ZND-like equations.
%
%\item Get a method for calculating these pathological, blowout detonations in 1D and see how they compare to FLASH
%\end{itemize}

\bibliography{townsley_snia}

\end{document}